\documentclass[12pt]{article}

\usepackage{epsfig}

\begin{document}
\hoffset=-1.0in
\voffset=-0.7in

\title{Astrophysical Evidence for the Non-Hermitian but $PT$-symmetric Hamiltonian of Conformal Gravity}

\author{Philip~D.~Mannheim \\ Department of Physics, University of Connecticut, Storrs, CT
06269, USA
\\ email: philip.mannheim@uconn.edu}

\date{May 25, 2012}

\maketitle

In this review we discuss the connection between two seemingly disparate topics, macroscopic gravity on astrophysical scales and Hamiltonians that are not Hermitian but $PT$ symmetric  on microscopic ones. In particular we show that the quantum-mechanical unitarity problem of the fourth-order derivative conformal gravity theory is resolved by recognizing that the scalar product appropriate to the theory is not the Dirac norm associated with a Hermitian Hamiltonian but is instead the norm associated with a non-Hermitian but $PT$-symmetric Hamiltonian. Moreover, the fourth-order theory Hamiltonian is not only not Hermitian, it is not even diagonalizable, being of Jordan-block form. With $PT$ symmetry we establish that conformal gravity is consistent at the quantum-mechanical level. In consequence, we can apply the theory to data, to find that the theory is capable of naturally accounting for the systematics of the rotation curves of a large and varied sample of 138 spiral galaxies without any need for dark matter. The success of the fits provides evidence for the relevance of non-diagonalizable but $PT$-symmetric Hamiltonians to physics.

\section{Introduction and Overview}
On January 19, 2000 a paper by myself and Aharon Davidson entitled ``Fourth order theories without ghosts" appeared on the arXiv \cite{Mannheim2000}. The abstract of the 2000 paper read:

``Using the Dirac constraint method we show that the pure fourth-order Pais-Uhlenbeck oscillator model is free of observable negative norm states. Even
though such ghosts do appear when the fourth order theory is coupled to a
second order one, the limit in which the second order action is switched off is
found to be a highly singular one in which these states move off shell. Given
this result, construction of a fully unitary, renormalizable, gravitational theory
based on a purely fourth order action in 4 dimensions now appears feasible."

\noindent
Also in the 2000 paper we identified a particular Hamiltonian $H(\epsilon=0)=8\gamma \omega^4 (2b^{\dagger}b+a^{\dagger}b 
+ b^{\dagger}a)+\omega$ and noted of it: 
``we see that the complete spectrum of eigenstates of
$H(\epsilon=0))$ is the set of all states $(b^{\dagger})^n|\Omega\rangle$,
a spectrum whose dimensionality is that of a one rather than a
two-dimensional harmonic oscillator, even while the complete Fock space
has the dimensionality of the two-dimensional oscillator.''

To be specific, what we had found in this paper and its follow-ups  \cite{Mannheim2005,Mannheim2007} was that  in the one-particle sector the Hamiltonian of the pure fourth-order Pais-Uhlenbeck oscillator model took the generic form 
\begin{eqnarray}
M=\pmatrix{1&1\cr 0&1},
\label{Z1}
\end{eqnarray}
and that even though the secular equation $|M-\lambda I|=0$ had two solutions (both equal to $1$), $M$  only had one eigenvector since the eigenvalue equation
\begin{eqnarray}
\pmatrix{1&1 \cr 0&1}\pmatrix{p \cr q}=\pmatrix{p+q\cr q}=\pmatrix{p
\cr q}
\label{Z2}
\end{eqnarray}
only had one eigenvector solution, viz. the $q=0$ one that obeyed
\begin{eqnarray}
\pmatrix{1&1 \cr 0&1}\pmatrix{1 \cr 0}=\pmatrix{1\cr 0}.
\label{Z3}
\end{eqnarray}
The one-particle sector Hamiltonian was thus of a non-diagonalizable, Jordan-block form. And with its missing an eigenvector, its eigenvector basis was incomplete (in consequence of which the two energy eigenvalues had to be equal since they had to share a common eigenvector), and $M$ could not be diagonalized by a similarity transformation.  Even though the Hamiltonian was not Hermitian, nonetheless all of its eigenvalues (as well as those in the multi-particle sectors of the theory) were real. The Pais-Uhlenbeck Hamiltonian thus serves as an example of a non-Hermitian Hamiltonian that has all eigenvalues real, to thus show that while Hermiticity of an operator is sufficient to yield real eigenvalues, it is not  necessary.

Aharon Davidson and I had been exploring the Pais-Uhlenbeck oscillator model because it serves as a prototype for the fourth-order conformal gravity theory that I had been advancing (see e.g. \cite{Mannheim2006,Mannheim2012a}) as a candidate alternative to standard gravity and string theory. My interest in the conformal gravity theory was triggered by the fact that in a conformal invariant theory the cosmological constant would have to be zero, with conformal gravity thus serving as a good starting point for addressing the cosmological constant problem. Problematically though, when quantized, fourth-order theories such as conformal gravity were thought to possess negative norm states and not be unitary. However, through our work Aharon Davidson and I were able to show that the theory could solve its ghost problem by having the ghosts not appear as on-shell energy eigenstates, to thereby potentially open the road to the construction of a consistent quantum gravitational theory in four spacetime dimensions.

As such, the pure fourth-order Pais-Uhlenbeck oscillator model provides an example of a $PT$-symmetric Hamiltonian at a so-called exceptional point -- a point at which the energy eigenspectrum is incomplete and the Hamiltonian cannot be brought to a diagonal form by a similarity transformation. However, back in 2000 I had no idea that the theory was a $PT$ theory, let alone one at an exceptional point, and only learned of it from Carl Bender in 2007 when the two of us started to collaborate on the theory  \cite{Bender2008a,Bender2008b,Bender2008c}. With Carl Bender and his collaborators only having started to develop the concept of non-Hermitian but $PT$-symmetric Hamiltonians in 1998 by showing that such Hamiltonians could have real eigenvalues (see e.g. \cite{Bender2007}), my 2000 paper with Aharon Davidson thus provides one of the first examples of a theory with a  non-Hermitian but $PT$-symmetric Hamiltonian at an exceptional point.

As first presented by Pais and Uhlenbeck \cite{Pais1950}, the Pais-Uhlenbeck model is based on an action
\begin{equation}
I_{\rm PU}={\gamma\over 2}\int dt[ \ddot{z}^2-(\omega_1^2+\omega_2^2)\dot{z}^2+\omega_1^2\omega_2^2z^2]
\label{Z4}
\end{equation}
that describes the dynamics of a single, non-relativistic coordinate $z(t)$. The associated Euler-Lagrange equation of motion is of the fourth-order derivative form 
\begin{equation}
\left[{d^4\over dt^4}+(\omega_1^2+\omega_2^2){d^2\over dt^2}+\omega_1^2\omega_2^2\right]z(t)
=\left[{d^2\over dt^2}+\omega_1^2\right]\left[{d^2\over dt^2}+\omega_2^2\right]z(t)=0,
\label{Z5}
\end{equation}
and the associated propagator is of the form
\begin{equation}
D_{\rm PU}(\omega^2)=\frac{1}{(\omega^2-\omega_1^2)(\omega^2-\omega_2^2)}
=\frac{1}{(\omega_1^2-\omega_2^2)}\left[ \frac{1}{(\omega^2-\omega_1^2)}- \frac{1}{(\omega^2-\omega_2^2)}\right],
\label{Z6}
\end{equation}
viz. a propagator whose poles are entirely on the real energy axis.

Two things in the structure of the propagator $D_{\rm PU}(\omega^2)$ immediately stand out: the presence of the relative minus sign in the unequal-frequency partial fraction decomposition and the singularity of the prefactor in this decomposition in the equal-frequency limit. We thus recognize two distinct sectors of the theory, the non-singular case of unequal frequencies  ($\omega_1\neq\omega_2$) and the singular case of equal frequencies ($\omega_1=\omega_2$). Prior to my work with Carl Bender the presence of the relative minus sign in the unequal-frequency propagator had always been taken to indicate that some of the states in the theory would have to have negative norm (since some of the poles would have negative residues), with the theory then not being unitary. And the result obtained by myself and  Aharon Davidson was that even if there were such negative norm states in the unequal-frequency theory, because of the singular nature of the equal-frequency limit, in the limit the Hamiltonian would lose some of eigenstates, and none of the energy eigenstates that remained would have negative norm. 

My collaboration with Carl Bender then went further by establishing that the unequal-frequency theory itself had a realization that was unitary, and that it too could be reinterpreted as a $PT$ rather than a Hermitian theory, and that then even in the unequal-frequency case there would be no states of negative norm.  Thus for both the unequal-frequency and equal-frequency theories the Hamiltonian would not be Hermitian. However, since all the poles of the $D_{\rm PU}(\omega^2)$ propagator lie on the real axis, all energy eigenvalues would be real, with the Hamiltonian then being $PT$ symmetric. At the time Carl Bender and I first studied the Pais-Uhlenbeck model we showed that the Hamiltonian was $PT$ symmetric simply by examining the behavior of the fields in it under a $PT$ transformation. Subsequently, we established  a general theorem \cite{Bender2010} that stated that if a Hamiltonian did not have a $PT$ symmetry (as understood in general to mean the product of discrete linear and discrete anti-linear operators), some of its eigenvalues would have to be complex, with no non-$PT$-symmetric Hamiltonian then being able to have a set of eigenvalues that was entirely real. Since the energy eigenvalues of the Pais-Uhlenbeck theory were all real and the Hamiltonian was not Hermitian, the theory thus had to be a $PT$ theory.

While my own interest had primarily been in the equal-frequency theory since it is a non-relativistic prototype of pure fourth-order gravity, the unequal-frequency theory is of interest not only because it is the quantum-mechanical limit of the often studied second-order plus fourth-order gravity theory, but also because it allows one to monitor what precisely happens in the limit in which the second-order piece is switched off. If one ignores metric indices (which as we show below we can), the issues involved can be illustrated by consideration of the second-order plus fourth-order scalar field theory with action, equation of motion and propagator of the form
\begin{eqnarray}
I_{\rm S}=\frac{1}{2}\int d^4x\left[\partial_{\mu}\partial_{\nu}\phi\partial^{\mu}
\partial^{\nu}\phi-M^2\partial_{\mu}\phi\partial^{\mu}\phi\right],
\label{Z7}
\end{eqnarray}
\begin{equation}
(\partial_t^2-\nabla^2)(\partial_t^2-\nabla^2+M^2)\phi(\bar{x},t)=0, 
\label{Z8}
\end{equation}
\begin{eqnarray}
D_{\rm S}(k^2)=\frac{1}{k^2(k^2+M^2)}
=\frac{1}{M^2}\left(\frac{1}{k^2}-\frac{1}{k^2+M^2}\right),
\label{Z9}
\end{eqnarray}
where $k^2=-k_0^2+\bar{k}^2$. For this propagator a standard Feynman contour integration yields a positive frequency contribution of the form 
\begin{equation}
D_{\rm S}(\bar{x},\bar{x}^{\prime},E)=\frac{1}{M^2}\left[\sum_n~\frac{\psi_n(\bar{x})\psi_n^*(\bar{x}^{\prime})}{E-E_n} -\sum_m~\frac{\psi_m(\bar{x})\psi_m^*(\bar{x}^{\prime})}{E-E_m}\right],
\label{Z10}
\end{equation}
where $\psi_n(\bar{x})=e^{i\bar{k}_n\cdot\bar{x}}$, $\psi_m(\bar{x})=e^{i\bar{k}_m\cdot\bar{x}}$, $E_n=|\bar{k}_n|$, $E_m=(\bar{k}_m^2+M^2)^{1/2}$. An identification of the propagator with the Green's function $\langle \Omega|T(\phi(x)\phi(x^{\prime}))|\Omega\rangle$ would then require the closure relation for the energy eigenstates to be of the form
\begin{equation}
\sum_n |n\rangle\langle n|- \sum_m |m\rangle\langle m|=I,
\label{Z11}
\end{equation}
with the presence of negative residues in the propagator translating into the presence of negative norm states in the Hilbert space. 

To avoid this outcome we must not identify the propagator with $\langle \Omega|T(\phi(x)\phi(x^{\prime}))|\Omega\rangle$ but must instead seek some other identification for it, one in which a closure relation involving negative norm states  would not then appear. That we might be able to do this at all is due to the fact that the propagator $D_{\rm S}(k^2)$ is a c-number quantity that is constructed from a c-number differential equation of the form $(\partial_t^2-\nabla^2)(\partial_t^2-\nabla^2+M^2)D_{\rm S}(x,x^{\prime})=\delta^4(x-x^{\prime})$, and as such it itself makes no reference to any Hilbert space. There is thus some freedom in relating the c-number propagator to an underlying q-number field theory, with other possible quantum-field-theoretic representations of the propagator being possible. And indeed, it is only after second-quantizing the scalar field in a very specific way, viz. the canonical $\phi(x)=\sum(a(\bar{k})e^{ik\cdot x}+a^{\dagger}(\bar{k})e^{-ik\cdot x})$ where the operator associated with the negative-frequency sector is expressly the Hermitian conjugate of the operator $a(\bar{k})$ associated with the positive-frequency sector, that we then construct the propagator as a matrix element of products of field operators in a vacuum state $|\Omega \rangle$ that $a(\bar{k})$ annihilates on the right  according to $a(\bar{k})|\Omega\rangle=0$ and $a^{\dagger}(\bar{k})$ annihilates on the left according to $\langle \Omega|a^{\dagger}(\bar{k})=0$. 

While this canonical prescription is completely standard in quantum field theory, it was established in field theories where the Hamiltonian and the fields were all Hermitian. Thus in the non-Hermitian but $PT$-symmetric case we can no longer identify the propagator this way. Rather, in the $PT$ case one must distinguish between left and right vacua. Specifically, if we expand the scalar field in creation and annihilation operators according to $\phi(x)=\sum(a(\bar{k})e^{ik\cdot x}+\hat{a}(\bar{k})e^{-ik\cdot x})$ where the scalar field is not now Hermitian and $\hat{a}(\bar{k})$ is not the Hermitian conjugate of $a(\bar{k})$, the left and right vacua will be  defined as the states that are annihilated according to $a(\bar{k})|\Omega_{\rm R}\rangle=0$, $\langle \Omega_{\rm L}|\hat{a}(\bar{k})=0$. In this case $\langle \Omega_{\rm L}|$ will not be the Hermitian conjugate of $|\Omega_{\rm R}\rangle$, and the propagator will be given not by the canonical $\langle \Omega|T(\phi(x)\phi(x^{\prime}))|\Omega\rangle$ but by $\langle \Omega_{\rm L}|T(\phi(x)\phi(x^{\prime}))|\Omega_{\rm R}\rangle$ instead. In the non-Hermitian but $PT$ case $\langle \Omega_{\rm L}|$ and $\langle \Omega_{\rm R}|$ while distinct are related according to $\langle \Omega_{\rm L}|=\langle \Omega_{\rm R}|V$, where $V$ is the operator that relates a non-Hermitian Hamiltonian $H$ to its adjoint according to $VHV^{-1}=H^{\dagger}$ (see e.g. \cite{Mannheim2012b} and references therein). In this case the closure relation and normalization conditions are of the form 
\begin{equation}
\sum_i |n_i\rangle\langle n_i|V=I,\qquad \langle n_i|V|n_j \rangle=\delta_{ij},
\label{Z12}
\end{equation}
and the propagator is given by $\langle \Omega_{\rm L}|T(\phi(x)\phi(x^{\prime}))|\Omega_{\rm R}\rangle =\langle \Omega_{\rm R}|VT(\phi(x)\phi(x^{\prime}))|\Omega_{\rm R}\rangle$.  With there being no states with $\langle n_i|V|n_j \rangle=-\delta_{ij}$, there are no states of negative norm, with it instead being the presence of the $V$ operator that generates the relative minus sign in the propagator. Specifically, the $V$ operator is related to the $C$ operator of $PT$ theory \cite{Bender2007}, an operator that commutes with the Hamiltonian and obeys $C^2=I$, with the relative minus sign in the propagator being due to the fact that the eigenvalues of the $C$ operator are $+1$ and $-1$. The relative minus sign is thus associated with the operators and not with any negative norm states. Thus what had led to the belief that the $D_{\rm S}(k^2)$ propagator had to be associated with negative norm states was the assumption that one had to identify the propagator as $\langle \Omega|T(\phi(x)\phi(x^{\prime}))|\Omega\rangle$ and had to use the Dirac norm. While such a Dirac norm is appropriate in the Hermitian case, as we see, the non-Hermitian situation is much richer, with the bra state in a scalar product  not in general needing to be the Hermitian conjugate of the ket.

As can be seen from the structure of the scalar field theory action $I_{\rm S}$ and propagator $D_{\rm S}(k^2)$, not only, in analog to the Pais-Uhlenbeck $D_{\rm PU}(\omega^2)$, do we encounter both a relative minus sign (when $M^2$ is non-zero) and a singularity (when we set $M^2=0$), in addition we see that in the $M^2\rightarrow 0$ limit the second-order term in the $I_{\rm S}$ action drops out, with the action becoming pure fourth order. Since the identification
\begin{equation}
\phi(\bar{x},t)\sim z(t) e^{i\bar{k}\cdot\bar{x}},~~\omega_1=(\bar{k}^2+M^2)^{1/2},~~\omega_2=|\bar{k}|
\label{Z13}
\end{equation}
(and thus $\omega_1^2-\omega_2^2 =M^2$) reduces the scalar field theory to the Pais-Uhlenbeck theory, we see that study of the Pais-Uhlenbeck theory enables us to study unitarity issues both in second- plus fourth-order derivative quantum field theories and in pure fourth-order ones. We thus turn now to a detailed study of the Pais-Uhlenbeck theory, and after establishing that fourth-order theories are unitary, we make an application of the pure fourth-order conformal gravity theory to galactic rotation curves.

\section{The Pais-Uhlenbeck Oscillator Model} 

As introduced, the Pais-Uhlenbeck action $I_{\rm PU}$ contains three degrees of freedom, $z$, $\dot{z}$, and $\ddot{z}$. This is too many degrees of freedom for one oscillator but not enough for two. Consequently the theory is constrained. To handle the constraints we replace $\dot{z}$ by a new variable $x$, and using the method of Dirac constraints obtain \cite{Mannheim2000,Mannheim2005} a Hamiltonian of the form 
\begin{equation}
H_{\rm PU}=\frac{p_x^2}{2\gamma}+p_zx+\frac{\gamma}{2}\left(\omega_1^2+\omega_2^2
\right)x^2-\frac{\gamma}{2}\omega_1^2\omega_2^2z^2,
\label{Z14}
\end{equation}
with two canonical pairs that obey 
\begin{equation}
\left[x,p_x\right]=i,~~~~\left[z,p_z\right]=i.
\label{Z15}
\end{equation}
If we proceed canonically we can introduce two sets of creation and annihilation operators 
\begin{eqnarray}
z&=&a_1e^{-i\omega_1t}+a_1^{\dagger}e^{i\omega_1t}+a_2e^{-i\omega_2t}+a_2^{\dagger}e^{i\omega_2t},
\nonumber \\
p_z&=&i\gamma\omega_1\omega_2^2
(a_1e^{-i\omega_1t}-a_1^{\dagger}e^{i\omega_1t})+i\gamma\omega_1^2\omega_2(a_2e^{-i\omega_2t}-a_2^{\dagger}e^{i\omega_2t}),
\nonumber\\
x&=&-i\omega_1(a_1e^{-i\omega_1t}-a_1^{\dagger}e^{i\omega_1t})-i\omega_2(a_2e^{-i\omega_2t}-a_2^{\dagger}e^{i\omega_2t}),
\nonumber \\
p_x&=&-\gamma\omega_1^2 (a_1e^{-i\omega_1t}+a_1^{\dagger}e^{i\omega_1t})-\gamma\omega_2^2(a_2e^{-i\omega_2t}+a_2^{\dagger}e^{i\omega_2t})
\label{Z16}
\end{eqnarray}
to find that the Hamiltonian and commutation relations take the form 
\begin{equation}
H_{\rm PU}=2\gamma(\omega_1^2-\omega_2^2)(\omega_1^2 a_1^\dag a_1-\omega_2^2a_2^\dag a_2)
+(\omega_1+\omega_2)/2, 
\label{Z17}
\end{equation}
\begin{equation}
[a_1,a_1^\dag]=\frac{1}{2\gamma\omega_1
\left(\omega_1^2-\omega_2^2\right)},\qquad
[a_2,a_2^\dag]=-\frac{1}{2\gamma\omega_2
\left(\omega_1^2-\omega_2^2\right)}.
\label{Z18}
\end{equation}

While we have thus succeeded in diagonalizing the unequal-frequency Hamiltonian via second quantization, we see that the limit of equal frequencies is singular. We will thus need to treat the equal-frequency case separately, and shall thus discuss the unequal-frequency case first. For the unequal-frequency case we note the relative minus sign between the two commutators. Thus in order to avoid negative norm states we must unconventionally define the vacuum as the one that $a_1$ and $a_2^{\dagger}$ annihilate (we take $\omega_1>\omega_2$ for definitiveness), with the energy and norm of the vacuum then being given by
\begin{eqnarray}
a_1|\Omega\rangle=a_2^{\dagger}|\Omega\rangle=0,~~~\langle \Omega |a_2^{\dagger}a_2|\Omega\rangle=\frac{\langle \Omega |\Omega\rangle}{2\gamma\omega_2
\left(\omega_1^2-\omega_2^2\right)}
>0,~~~H_{\rm PU}|\Omega\rangle=\frac{1}{2}(\omega_1-\omega_2)|\Omega\rangle.
\label{Z19}
\end{eqnarray}
In such a vacuum states of the form $(a_2)^n|\Omega \rangle$ have negative energy, with the energy spectrum being unbounded from below, just as  is to be anticipated given the presence of the $-(\gamma/2)\omega_1^2\omega_2^2z^2$ term in $H_{\rm PU}$.

To avoid such an outcome we could instead work in a Hilbert space in which $a_1$ and $a_2$ annihilate the vacuum. However, then we obtain
\begin{eqnarray}
a_1|\Omega\rangle=a_2|\Omega\rangle=0,~~~\langle\Omega |a_2a_2^{\dagger}|\Omega\rangle=-
\frac{\langle \Omega |\Omega\rangle}
{2\gamma\omega_2
\left(\omega_1^2-\omega_2^2\right)}
<0,~~~H_{\rm PU}|\Omega\rangle=\frac{1}{2}(\omega_1+\omega_2)|\Omega\rangle.
\label{Z20}
\end{eqnarray}
While we no longer have any states with negative energy, we instead now have states with negative norm.

As we see, both of the above realizations of the commutation algebra are undesirable and the problem even appears to be insurmountable. However, quantum mechanics is a global theory, and we must thus take global considerations into account. To this end we make a 
Schr\"odinger wave mechanics representation of the commutation algebra of the form 
\begin{equation}
p_x=-i\frac{\partial}{ \partial x},\qquad p_z=-i\frac{\partial}{ \partial z},
\label{Z21}
\end{equation}
to obtain a stationary state with energy $(\omega_1+\omega_2)/2$, viz. the ground state in the realization of the theory where energies are bounded from below, with its wave function being given by $\psi_0(z,x)e^{-i(\omega_1+\omega_2)t/2}$ where
\begin{equation}
\psi_0(z,x)=\exp\bigg[\frac{\gamma}{2}(\omega_1+\omega_2)
\omega_1\omega_2z^2
+i\gamma\omega_1\omega_2zx -\frac{\gamma}{2}(\omega_1+\omega_2)x^2
\bigg].
\label{Z22}
\end{equation}
As we see, while this wave function has good asymptotic behavior at large $x$, it diverges at large $z$, and is thus not normalizable.  Since we can only represent the momentum operator as $-i\partial_z$ when it acts on wave functions that are well-behaved at large $z$, we see that this representation for $p_z$ is invalid. In an integration by parts of $\int dz \psi^*(z)(-i\partial_z)\psi(z)$ we would not be able to throw away surface terms, and thus have to conclude that in the bounded from below energy sector the operator $p_z$ cannot be Hermitian. Similarly, in this sector the vacuum normalization evaluates to $\langle \Omega |\Omega\rangle=\langle \Omega |\int dzdx|z,x\rangle\langle z,x|\Omega\rangle=\int dzdx \psi^*_0(z,x)\psi_0(z,x)$, to thus be infinite. Thus the global information that is missing from (\ref{Z20}) is that we cannot normalize $\langle \Omega |\Omega\rangle$ to a finite value, and that accordingly no closure relation of the form given in (\ref{Z11}) can hold.

While $\psi_0(z,x)$ is not normalizable when $z$ is real, we note that it would be nicely bounded if $z$ were pure imaginary. Hence the theory will be well-behaved if take $p_z$ not to be Hermitian but to be anti-Hermitian instead. With the $z$ operator then being anti-Hermitian too, the troublesome  $-(\gamma/2)\omega_1^2\omega_2^2z^2$ term now is bounded from below. Since we can continue to maintain Hermiticity for $x$ and $p_x$, we see that because of the $p_zx$ cross term $H_{\rm PU}$ is not Hermitian. However, its eigenvalues are real, so $H_{\rm PU}$ emerges as a $PT$-symmetric Hamiltonian. Since we had made the creation and annihilation operator expansion of (\ref{Z16}) on the assumption that $z$ and $p_z$ were Hermitian, we see that (\ref{Z16}) is not a valid expansion in the realization of the theory in which the energy is bounded from below. Since use of (\ref{Z16})  led us to states with negative norm, we must seek an alternate formulation of the theory, one  that uses some entirely different norm.

To find such a norm it is convenient to work not with anti-Hermitian $z$ and $p_z$ but to instead  introduce Hermitian operators $y$ and $q$ according to
\begin{equation}
y=e^{\pi p_zz/2}ze^{-\pi p_zz/2}=-iz,\qquad q=e^{\pi p_zz/2}p_ze^{-\pi p_zz/2}=ip_z.
\label{Z23}
\end{equation}
These operators also form a canonical pair with $[y,q]=i$, with the transformation in (\ref{Z23}) being one of a broad class of transformations that leave quantum commutators (and equally classical Poisson brackets) unchanged. Specifically, a commutator such as $[z,-i\partial_z]$ is left unchanged if  $z$ is transformed to  $e^{i\theta}z$  since  $\partial_z$ is then transformed to $e^{-i\theta}\partial_z$. In and of itself such a transformation would be without content unless in varying $\theta$ we can transit between regions in the complex $z$ plane with differing asymptotic behavior for the wave functions. The complex $z$ plane thus breaks up into regions known as Stokes wedges \cite{Bender2007} with differing asymptotic behaviors, and one is able to represent the $[z,p_z]=i$ commutator in the differential form $[z,-i\partial_z]$ only when it acts on states $\psi(e^{i\theta} z)$ in a domain in the complex $z$ plane where $\psi(e^{i\theta}z)$ is asymptotically bounded.

In terms of the operators $y$ and $q$ and an $x$ and a $p_x=p$ that obey $[x,p]=i$, the Hamiltonian takes the non-Hermitian form 
\begin{equation}
H=\frac{p^2}{2\gamma}-iqx+\frac{\gamma}{2}\left(\omega_1^2+\omega_2^2
\right)x^2+\frac{\gamma}{2}\omega_1^2\omega_2^2y^2,
\label{Z24}
\end{equation}
and its $PT$ symmetry is made manifest via the assignments:  $p$ and $x$ are odd under $P$, $q$ and $y$ are $P$ even, $p$ and $y$ are $T$ odd, $q$ and $x$ are $T$ even. Finally, using the techniques of $PT$ theory we introduce a Hermitian operator $Q$ defined as \cite{Bender2008a}
\begin{equation}
Q=\alpha [pq+\gamma^2\omega_1^2\omega_2^2 xy],\qquad \alpha=\frac{1}{\gamma\omega_1\omega_2}{\rm log}\left(\frac{\omega_1+\omega_2}{\omega_1-\omega_2}\right),
\label{Z25}
\end{equation}
to find that the unequal-frequency Hamiltonian $\tilde{H}$ given as 
\begin{equation}
{\tilde H}=e^{-{Q}/2}H_{\rm PU}e^{{Q}/2}
=\frac{p^2}{2\gamma}+\frac{q^2}{2\gamma\omega_1^2}+
\frac{\gamma}{2}\omega_1^2x^2+\frac{\gamma}{2}\omega_1^2\omega_2^2y^2
\label{Z26}
\end{equation}
is Hermitian and diagonal, with all of its energy eigenvalues being  real and positive. The unequal-frequency $H_{\rm PU}$ while not itself Hermitian can be brought to a Hermitian form by a similarity transform. 

Since Hermitian theories possess no states with negative norm, it follows that the unequal-frequency $H_{\rm PU}$ theory must possess none either. To see what form the $H_{\rm PU}$ theory norms do take, we note that since $Q$ is Hermitian rather than anti-Hermitian, the unequal frequency $H_{\rm PU}$ has been diagonalized by a similarity transformation that is non-unitary. Since similarity transformations do not preserve orthonormality of basis vectors, the eigenstates of $H$ and $\tilde{H}$ are not unitarily equivalent. Rather, if we define eigenstates of $H_{\rm PU}$ and $\tilde{H}$ according to $H_{\rm PU}|n\rangle=E_n|n\rangle$ and ${\tilde H}|{\tilde n}\rangle=E_n|{\tilde n}\rangle$, we find that these states and their conjugates are related according to
\begin{eqnarray}
&&{\tilde H}|{\tilde n}\rangle=E_n|{\tilde n}\rangle,\qquad
H_{\rm PU}|n\rangle=E_n|n\rangle,\qquad
|n\rangle=e^{Q/2}|{\tilde n}\rangle
\nonumber\\
&&\langle{\tilde n}|{\tilde H}=E_n\langle{\tilde n}|,\qquad
\langle n|\equiv\langle{\tilde n}|e^{Q/2},\qquad
\langle n|e^{-Q}H_{\rm PU}=\langle n|e^{-Q}E_n
\label{Z27}
\end{eqnarray}
with the Hermitian conjugate $\langle n|$ of the right-eigenstate  $|n\rangle$ of $H_{PU}$ not being a left-eigenstate of $H_{\rm PU}$. In terms of the operator $V$ we introduced earlier we can identify $V=e^{-Q}$, and can check directly that $VH_{\rm PU}V^{-1}=H^{\dagger}_{\rm PU}$ just as required. (In passing we note that in this case the $C$ operator of $PT$ theory can be written  in terms of $Q$ and the parity operator $P$ as $C=e^{Q}P$ \cite{Bender2008a}, with it obeying $C^2=I$ since $Q$ obeys $PQ=-QP$.) 
Finally, since $\tilde{H}$ is unitary, we obtain normalization and closure relations of the form 
\begin{eqnarray}
&&\langle \tilde{n}|\tilde{m}\rangle=\delta_{m,n},\qquad \Sigma | \tilde{n}\rangle\langle  \tilde{n}|=I,\qquad \tilde{H}=\Sigma | \tilde{n}\rangle E_n\langle  \tilde{n}|,
\nonumber\\
&&\langle n|e^{-Q}|m\rangle=\delta_{m,n},\qquad \Sigma |n\rangle\langle n|e^{-Q}=I,\qquad H_{\rm PU}=\Sigma |n\rangle E_n\langle n|e^{-Q}
\label{Z28}
\end{eqnarray}
with there being no states of negative norm, and with the similarity transformation thus taking us between a skew basis and an orthonormal one. In the $PT$ realization of the unequal-frequency theory then there are no states of negative energy and no states of negative norm,  and the theory is fully physically acceptable.

\section{The Singular Equal-Frequency Limit}

Inspection of (\ref{Z25}) shows that in equal-frequency limit the operator $Q$ becomes singular,  with the transformation to $\tilde{H}$ no longer being valid. To see what happens in the limit we consider the two one-particle states of the realization of the unequal frequency theory in which the energy is bounded from below, viz.  
\begin{eqnarray}
\psi_1(x,y,t)=(x+\omega_2y)\psi_0(x,y,t)e^{-i\omega_1t},\qquad E_1=E_0+\omega_1,
\nonumber \\
\psi_2(x,y,t)=(x+\omega_1y)\psi_0(x,y,t)e^{-i\omega_2t},\qquad E_2=E_0+\omega_2,
\label{Z29}
\end{eqnarray}
where $\psi_0(x,y,t)$ is the ground state wave function 
\begin{eqnarray}
\psi_0(x,y,t)&=&\exp\left[-\frac{\gamma}{2}(\omega_1+\omega_2)x^2-\gamma\omega_1\omega_2yx- 
\frac{\gamma}{2}(\omega_1+\omega_2)\omega_1\omega_2y^2
-iE_0t\right]
\label{Z30}
\end{eqnarray}
with energy $E_0=(\omega_1+\omega_2)/2$.
As we see, in the limit not only do the two one-particle energies become equal, the two wave functions become identical with the two one-particle states having collapsed into a single state. To confirm that the Hamiltonian has thus become a non-diagonalizable, Jordan-block Hamiltonian, we note that in the unequal-frequency one-particle sector the Hamiltonian takes the form \cite{Bender2008b}
\begin{eqnarray}
H_{\rm 1P}(\epsilon)=\frac{1}{2\omega}\pmatrix{4\omega^2+\epsilon^2&4\omega^2-
\epsilon^2\cr\epsilon^2&4\omega^2-\epsilon^2},
\label{Z31}
\end{eqnarray}
where we have set $\omega_1=\omega+\epsilon$, $\omega_2=\omega-\epsilon$. For this Hamiltonian the eigenstates are given by
\begin{eqnarray}
|2\omega+\epsilon\rangle \equiv \pmatrix{
2\omega-\epsilon
\cr\epsilon},\qquad 
|2\omega-\epsilon\rangle \equiv \pmatrix{2
\omega+\epsilon\cr-\epsilon}
\label{Z32}
\end{eqnarray}
corresponding to eigenenergies $2\omega+\epsilon$ and $2\omega-\epsilon$. 

If we introduce the operator
\begin{eqnarray}
S(\epsilon)&=&\frac{1}{2\epsilon\omega^{1/2}(2\omega+\epsilon)^{1/2}}\pmatrix{2\omega+
\epsilon&-(4\omega^2-\epsilon^2)\epsilon\cr\epsilon&(2\omega+\epsilon)\epsilon^2
}
\label{Z33}
\end{eqnarray}
we find that $H_{\rm 1P}(\epsilon)$ is diagonalized by this $S(\epsilon)$ according to 
\begin{eqnarray}
S^{-1}(\epsilon)\left(\frac{1}{2\omega}\right)\pmatrix{4\omega^2+\epsilon^2&4\omega^2-
\epsilon^2\cr\epsilon^2&4\omega^2-\epsilon^2}S(\epsilon)
=\pmatrix{2\omega+\epsilon&0\cr 0&2\omega-\epsilon}
\label{Z34}
\end{eqnarray}
Then, in the $\epsilon \rightarrow 0$ limit, we see that $S(\epsilon)$ becomes undefined and  $H_{\rm 1P}(\epsilon)$ becomes the Jordan-block Hamiltonian
\begin{eqnarray}
H_{1P}(\epsilon=0)=2\omega\pmatrix{1&1\cr 0&1}
\label{Z35}
\end{eqnarray}
with two degenerate energies but only one eigenstate. (As noted in \cite{Bender2008b}, by setting  $y(\epsilon=0)=e^{-i\omega t}[-i(a-b)-2b\omega t]+e^{i\omega t}[-i(\hat{a}-\hat{b})+2\hat{b}\omega t]$  one can introduce creation and annihilation operators at $\epsilon =0$ that obey $[a,\hat{a}]=0$, $[b,\hat{b}]=0$, $[a,\hat{b}]=[b,\hat{a}]=1/8\gamma\omega^3$, to then obtain $H_{\rm PU}(\epsilon=0)=8\gamma \omega^4 (2\hat{b}b+\hat{a}b + \hat{b}a)+\omega$, with this form for  $H_{\rm PU}(\epsilon=0)$ updating the form given in \cite{Mannheim2000} since  $\hat{a}$ and $\hat{b}$ are not the Hermitian conjugates of $a$ and $b$. With this form for $H_{\rm PU}(\epsilon=0)$ the one-particle sector $H_{1P}(\epsilon=0)$ then follows.)

To ascertain what happened to the eigenstate of $H_{1P}(\epsilon)$ that went missing in the limit, we take the limit of the ground state and of two particular linear combinations of the two one-particle eigenfunctions of the unequal-frequency theory, to obtain
\begin{equation}
\hat{\psi}_0(x,y,t)= \lim_{\epsilon\to0} \psi_0(x,y,t)=\exp\left[-\gamma\omega^3y^2
-\gamma\omega^2yx -\gamma\omega x^2-i\omega t\right],
\label{Z36}
\end{equation}
\begin{eqnarray}
\hat{\psi}_1(x,y,t)&=&\lim_{\epsilon\to0}\frac{\psi_2(x,y,t)+ \psi_1(x,y,t)}{2}=(x+\omega y)\hat{\psi}_0(x,y,t)e^{-i\omega t},
\nonumber \\
\hat{\psi}_{1a}(x,y,t)&=&\lim_{\epsilon\to0}\frac{\psi_2(x,y,t)- \psi_1(x,y,t)}{2\epsilon}
=\left[(x+\omega
y)it+y\right]\hat{\psi}_0(x,y,t)e^{-i\omega t}.
\label{Z37}
\end{eqnarray}
As we see, the non-singular combination $\hat{\psi}_1(x,y,t)$ remains an eigenstate in the limit, but the singular combination $\hat{\psi}_{1a}(x,y,t)$ does not. However, while $\hat{\psi}_{1a}(x,y,t)$ is not an energy eigenstate, it is a non-stationary solution to the time-dependent 
Schr\"odinger equation
\begin{eqnarray}
i\frac{\partial}{\partial t}\hat{\psi}(x,y,t)=\left(-\frac{1}{2\gamma}
\frac{\partial^2}{\partial x^2}-x\frac{\partial}{\partial y}+\gamma\omega^2x^2+
\frac{\gamma}{2}\omega^4y^2\right)\hat{\psi}(x,y,t).
\label{Z38}
\end{eqnarray}
Despite the fact that the non-stationary states are not energy eigenstates,  the set of stationary plus non-stationary states combined is complete since together they provide just the right number of independent polynomial functions of $x$ and $y$ that are needed to form an arbitrary localized wave packet in the $(x,y)$ plane \cite{Bender2008b}. Moreover, as shown in \cite{Bender2008b}, the norm of any such packet is preserved in time, with time evolution  being unitary. As we thus see, in the event that a Hamiltonian is not diagonalizable, unitarity is still expressed as a completeness statement. However, now it is the set of solutions to the time-dependent plus time-independent Schr\"odinger equations combined that needs to be complete rather than solutions to the time-independent one alone.

\section{Conformal Gravity}

The conformal Weyl tensor defined as 
\begin{eqnarray}
C_{\lambda\mu\nu\kappa}&=&R_{\lambda\mu\nu\kappa}
+{1 \over 6}R^{\alpha}_{\phantom{\alpha}\alpha}\left[
g_{\lambda\nu}g_{\mu\kappa} -g_{\lambda\kappa}g_{\mu\nu}\right]
-{1 \over 2}\left[
g_{\lambda\nu}R_{\mu\kappa} -g_{\lambda\kappa}R_{\mu\nu}
-g_{\mu\nu}R_{\lambda\kappa} +g_{\mu\kappa}R_{\lambda\nu}\right]
\label{Z39}
\end{eqnarray}
is a quite remarkable geometric object. Under a local conformal transformation of the form
\begin{equation}
g_{\mu\nu}(x)\rightarrow e^{2\alpha(x)}g_{\mu\nu}(x)
\label{Z40}
\end{equation}
with arbitrary spacetime-dependent $\alpha(x)$ the Weyl tensor transforms as $C_{\lambda\mu\nu\kappa}\rightarrow e^{2\alpha(x)}C_{\lambda\mu\nu\kappa}$. Thus even though the Riemann tensor $R_{\lambda\mu\nu\kappa}$ and its $R_{\mu\kappa}=g^{\lambda\nu}R_{\lambda\mu\nu\kappa}$ and $R^{\alpha}_{\phantom{\alpha}\alpha}=g^{\mu\kappa}R_{\mu\kappa}$ Ricci contractions all acquire derivatives of $\alpha(x)$ under the conformal transformation (each being a second-derivative function of the metric), all of these derivative terms cancel identically in the Weyl tensor. As such, the Weyl tensor behaves under a conformal transformation the same way that the Maxwell tensor $F_{\mu\nu}$ behaves under a gauge transformation. It is thus very suggestive to endow gravity with such a conformal structure, so that as well as being general coordinate invariant, gravity will in addition possess local conformal invariance as well. With such additional symmetry the gravitational action is uniquely prescribed to be of the form  
\begin{equation}
I_{\rm W}=-\alpha_g\int d^4x (-g)^{1/2}C_{\lambda\mu\nu\kappa} C^{\lambda\mu\nu\kappa}
\equiv -2\alpha_g\int d^4x (-g)^{1/2}\left[R^{\mu\nu}R_{\mu\nu}-{1 \over 3}(R^{\alpha}_{\phantom{\alpha}\alpha})^2\right]
\label{Z41}
\end{equation}
where the gravitational coupling constant $\alpha_g$ is dimensionless, with no other terms being permitted. (To obtain a dimensionless $\alpha_g$ one needs four derivatives to compensate for the $\int d^4x$ measure, with the $C_{\lambda\mu\nu\kappa} C^{\lambda\mu\nu\kappa}$ term precisely providing the needed four derivatives). Associated with the $I_{\rm W}$ action are fourth-order derivative gravitational equations of motion of the form (see e.g. \cite{Mannheim2006})
\begin{equation}
4\alpha_g [2C^{\mu\lambda\nu\kappa}_{\phantom{\lambda\mu\nu\kappa};\lambda;\kappa}
-C^{\mu\lambda\nu\kappa}R_{\lambda\kappa}]=4\alpha_gW^{\mu\nu}
=T^{\mu\nu},
\label{Z42}
\end{equation}
where $W^{\mu\nu}$ is covariantly conserved and traceless and can be written in the form
\begin{eqnarray}
W^{\mu \nu}&=&
\frac{1}{2}g^{\mu\nu}(R^{\alpha}_{\phantom{\alpha}\alpha})   
^{;\beta}_{\phantom{;\beta};\beta}+
R^{\mu\nu;\beta}_{\phantom{\mu\nu;\beta};\beta}                     
 -R^{\mu\beta;\nu}_{\phantom{\mu\beta;\nu};\beta}                        
-R^{\nu \beta;\mu}_{\phantom{\nu \beta;\mu};\beta}                          
 - 2R^{\mu\beta}R^{\nu}_{\phantom{\nu}\beta}                                    
+\frac{1}{2}g^{\mu\nu}R_{\alpha\beta}R^{\alpha\beta}                                             
\nonumber \\
&-&\frac{2}{3}g^{\mu\nu}(R^{\alpha}_{\phantom{\alpha}\alpha})          
^{;\beta}_{\phantom{;\beta};\beta} 
+\frac{2}{3}(R^{\alpha}_{\phantom{\alpha}\alpha})^{;\mu;\nu}                           
+\frac{2}{3} R^{\alpha}_{\phantom{\alpha}\alpha}
R^{\mu\nu}                              
-\frac{1}{6}g^{\mu\nu}(R^{\alpha}_{\phantom{\alpha}\alpha})^2.
\label{Z43}
\end{eqnarray}                                 

As constructed, conformal gravity is thus a power-counting renormalizable theory of gravity since $\alpha_g$ is dimensionless. Moreover, conformal gravity is able to control the cosmological constant since conformal invariance does not permit the presence of any  $I=\int d^4x (-g)^{1/2}\Lambda$ term in the fundamental action. In addition, we note that since the cosmologically relevant  Robertson-Walker (RW) geometry is a conformally flat geometry (viz. one in which $C_{\lambda\mu\nu\kappa}=0$), in conformal cosmology (\ref{Z42}) reduces to 
\begin{equation}
T^{\mu\nu}=0,
\label{Z44}
\end{equation}
so unlike the (non-conformal) double-well Higgs potential, conformal gravity knows exactly where the zero of energy is, a feature of the theory that is central to solving the cosmological constant problem \cite{Mannheim2011a,Mannheim2012a}.

As well as exclude any cosmological constant term from the fundamental action, the same conformal symmetry also excludes the presence of any Einstein-Hilbert term $I_{\rm EH}=-(c^3/16 \pi G)\int d^4x(-g)^{1/2} R^{\alpha}_{\phantom{\alpha}\alpha}$ in the fundamental action. Despite this, the theory still admits of the Schwarzschild solution \cite{Mannheim1989,Mannheim1994} since $R_{\mu\nu}=0$ is an exact  solution to (\ref{Z42}) when $T_{\mu\nu}=0$. The conformal theory thus recovers both Newton's Law of Gravity and its familiar general-relativistic corrections (gravitational bending of light, gravitational redshift, and modifications to planetary orbits). As noted in \cite{Mannheim2006}, Einstein gravity is thus only sufficient to give the standard Schwarzschild metric phenomenology but not necessary.

Now while the vanishing of $R^{\mu\nu}$ implies the vanishing of $W^{\mu\nu}$, the converse is not true since the vanishing of $W^{\mu\nu}$ can be achieved without $R^{\mu\nu}$ vanishing. The vacuum vanishing of $W^{\mu\nu}$ thus has other solutions \cite{Mannheim1989,Mannheim1994}, solutions that can thus enable us  to distinguish between Einstein gravity and conformal gravity. As we shall see below, these other solutions are important on galactic distance scales where they enable the conformal theory to account for galactic rotation curve systematics without the need for dark matter.

To identify the quantum structure of the conformal theory, it is convenient to linearize the theory around flat spacetime according to $g_{\mu\nu}=\eta_{\mu\nu}+h_{\mu\nu}$ where $\eta_{\mu\nu}$ is the Minkowski metric and $h_{\mu\nu}$ is the fluctuation. Through second order in $h_{\mu\nu}$ it is found \cite{Mannheim2012c} that  (\ref{Z41}) and (\ref{Z42}) only depend on the traceless combination $K_{\mu\nu}=h_{\mu\nu}-(1/4)\eta_{\mu\nu}\eta^{\alpha\beta}h_{\alpha\beta}$. The equations of motion are found to simplify considerably in the transverse gauge where $\partial_{\mu}K^{\mu\nu}=0$, with the first order modification to $W^{\mu\nu}$ being given by 
\begin{equation}
W^{\mu\nu}(1)=\frac{1}{2}(\partial_{\alpha}\partial^{\alpha})^2 K^{\mu\nu}
\label{Z45}
\end{equation}
(with both sides of (\ref{Z45}) being traceless), and with the second-order fluctuation term in the conformal action $I_{\rm W}$ being given by
\begin{equation}
I_{\rm W}(2)=-\frac{\alpha_g}{2}\int d^4x \partial_{\alpha}\partial^{\alpha} K_{\mu\nu}\partial_{\beta}\partial^{\beta} K^{\mu\nu}.
\label{Z46}
\end{equation}
As we see, both (\ref{Z45}) and (\ref{Z46}) are diagonal in the $(\mu,\nu)$ indices, and so each component of $K_{\mu\nu}$ can be treated independently. Comparing with (\ref{Z7}) and (\ref{Z8}), we see that each independent component of $K_{\mu\nu}$ obeys the same dynamics as a fourth-order scalar field with $M^2=0$. The unitarity analysis of the scalar field theory this caries over identically to the conformal theory \cite{Mannheim2012a}. Thus despite the fourth-order nature of its equations of motion, the unitarity of microscopic conformal gravity is secured. Since the macroscopic limit of the conformal theory is obtained by taking matrix elements of its quantum field operators in states with an indefinite number of gravitational quanta, the consistency of conformal gravity as a macroscopic theory is thus secured as well, and so we turn now to a study of some of its macroscopic implications.

\section{Fitting Galactic Rotation Curves}

To fit galactic rotation curves one needs to determine the geometry for a static, spherically symmetric system in the conformal case. To this end it was noted in \cite{Mannheim1989} that though use of the underlying conformal symmetry the most general metric in the static, spherically symmetric case can be brought to the form
\begin{equation}
ds^2=B(r)dt^2-\frac{dr^2}{B(r)}-r^2d\theta^2-r^2\sin^2\theta d\phi^2,
\label{Z47}
\end{equation}
to thus be described in  terms of just the one function $B(r)$ where $r$ is the radial coordinate. Moreover, in \cite{Mannheim1994} it was shown that in metric given in (\ref{Z47})  the gravitational equation of motion given in (\ref{Z42}) takes the form of a fourth-order Poisson equation
\begin{equation}
\frac{3}{B(r)}(W^{0}_{\phantom{0}0}-W^{r}_{\phantom{r}r})=B^{\prime\prime\prime\prime}+\frac{4B^{\prime\prime\prime}}{r}=\nabla^4B=\frac{3}{4\alpha_gB(r)}(T^{0}_{\phantom{0}0}-T^{r}_{\phantom{r}r})=f(r).
\label{Z48}
\end{equation}
As derived, (\ref{Z48}) is exact and without approximation, showing that despite its complexity, in the presence of sufficient symmetry (\ref{Z42}) can be quite tractable. Moreover, for a general source function $f(r)$ (\ref{Z48}) can be integrated exactly to yield
\begin{eqnarray}
B(r)= -\frac{r}{2}\int_0^r
dr^{\prime}r^{\prime 2}f(r^{\prime})
-\frac{1}{6r}\int_0^r
dr^{\prime}r^{\prime 4}f(r^{\prime})
-\frac{1}{2}\int_r^{\infty}
dr^{\prime}r^{\prime 3}f(r^{\prime})
-\frac{r^2}{6}\int_r^{\infty}
dr^{\prime}r^{\prime }f(r^{\prime})+\hat{B}(r),
\label{Z49}
\end{eqnarray}                                 
\begin{eqnarray}
\frac{dB(r)}{dr}= -\frac{1}{2}\int_0^r
dr^{\prime}r^{\prime 2}f(r^{\prime})
+\frac{1}{6r^2}\int_0^r
dr^{\prime}r^{\prime 4}f(r^{\prime})
-\frac{r}{3}\int_r^{\infty}
dr^{\prime}r^{\prime }f(r^{\prime})+\hat{B}^{\prime}(r),
\label{Z50}
\end{eqnarray}                                 
where $\hat{B}(r)$ satisfies $\nabla^4\hat{B}=0$. 

In contrast to (\ref{Z49}) and (\ref{Z50}), we recall that for the second-order Poisson equation $\nabla^2\phi(r)=g(r)$, the solution is of the form
\begin{equation}
\phi(r)= -\frac{1}{r}\int_0^r
dr^{\prime}r^{\prime 2}g(r^{\prime})-\int_r^{\infty}
dr^{\prime}r^{\prime }g(r^{\prime}),
\label{Z51}
\end{equation}                                 
\begin{equation}
\frac{d\phi(r)}{dr}= \frac{1}{r^2}\int_0^r
dr^{\prime}r^{\prime 2}g(r^{\prime}).
\label{Z52} 
\end{equation}                                 
Comparing (\ref{Z50}) and (\ref{Z52}), and noting the presence of an integral that continues all the way to infinity in (\ref{Z50}), we see that the novel feature of the fourth-order theory is that the gravitational force at radius $r$ receives contributions not just from local sources in the $0<r^{\prime}<r$ region but from global sources in the $r<r^{\prime}<\infty$ region as well. Thus unlike Newtonian gravity where local gravity is produced solely by local sources, in the fourth-order case global sources can contribute as well. Newtonian gravity is thus intrinsically local, while conformal gravity is intrinsically global. Since in Newtonian gravity one can only explain local gravitational observations though the use of local sources, a failure to account for observations using luminous sources alone leads to the need to postulate the presence of dark matter within any given system of interest. However, in the conformal theory there are additional contributions to the gravitational field within systems due to matter that is outside of them. As shown in \cite{Mannheim1997,Mannheim2011b,Mannheim2012d,O'Brien2012}, these global effects can remove the need for galactic dark matter, with dark matter potentially being nothing more than an attempt to describe global physics effects in purely local terms. To determine the effect of conformal gravity on galactic motions we thus need to determine the local contribution due to the material within each given galaxy and the global contribution due to the material in the rest of the Universe.

As regards first the local contribution, we note  that if in (\ref{Z49}) we consider a purely localized source of radius $r_0$, then on setting $\hat{B}(r)=1$, in the region $r>r_0$ we obtain
\begin{equation}
B(r>r_0)=1-\frac{2\beta}{r}+\gamma r,
\label{Z53}
\end{equation}
where
\begin{equation}
\gamma= -\frac{1}{2}\int_0^{r_0}dr^{\prime}r^{\prime 2}f(r^{\prime}),\qquad 
2\beta=\frac{1}{6}\int_0^{r_0}dr^{\prime}r^{\prime 4}f(r^{\prime}).
\label{Z54}
\end{equation}
In (\ref{Z53}) we obtain a Newtonian potential term even though there is no reference to any second-order Poisson equation. Thus, as noted earlier,  standard second-order gravity is only sufficient to give the Newtonian potential but not necessary. In addition we obtain a linear potential term. This term modifies standard gravity at large distances but not at small ones, i.e. it modifies gravity in precisely the region where the standard theory has to resort to dark matter.

In conformal gravity a local source such as a star produces a potential of the form
\begin{equation}
V^{*}(r)=\frac{(B(r)-1)c^2}{2}=-\frac{\beta^{*}c^2}{r}+\frac{\gamma^{*} c^2 r}{2}
\label{Z55}
\end{equation}
for each solar unit of material. For spiral galaxies the luminous material is typically distributed in a disk with surface brightness $\Sigma (R)=\Sigma_0e^{-R/R_0}$ at radial coordinate $R$. Integrating the potential of (\ref{Z55}) over such a distribution yields a contribution to circular velocities  in the plane of the disk of the form \cite{Mannheim2006}
\begin{eqnarray}
v_{{\rm LOC}}^2(R)&=&
\frac{N^*\beta^*c^2 R^2}{2R_0^3}\left[I_0\left(\frac{R}{2R_0}
\right)K_0\left(\frac{R}{2R_0}\right)-
I_1\left(\frac{R}{2R_0}\right)
K_1\left(\frac{R}{2R_0}\right)\right]
\nonumber \\
&+&\frac{N^*\gamma^* c^2R^2}{2R_0}I_1\left(\frac{R}{2R_0}\right)
K_1\left(\frac{R}{2R_0}\right),
\label{Z56}
\end{eqnarray} 
where $N^*=2\pi (M/L)\Sigma_0 R_0^2/M_{\odot}$ is the total number of stars in a galaxy, $M/L$ is the galactic mass to light ratio, and $M=N^*M_{\odot}$ is the visible galactic mass.

As regards the global contribution, we note that it consists of two components, a contribution from the homogeneous background Hubble flow, and a contribution due to inhomogeneities in it such as clusters of galaxies.  Since $W_{\mu\nu}$ vanishes identically in an RW geometry, the Hubble flow only contributes to the $\hat{B}(r)$ term in (\ref{Z49}). To determine the effect of the Hubble flow on motions within galaxies we need to transform from comoving Hubble flow coordinates to coordinates at rest in a galaxy. With the transformation 
\begin{equation}
\rho=\frac{4r}{2(1+\gamma_0r)^{1/2}+2 +\gamma_0 r},\qquad \tau=\int dt R(t)
\label{Z57}
\end{equation}                                 
effecting
\begin{eqnarray}
(1+\gamma_0r)dt^2-\frac{dr^2}{(1+\gamma_0r)}-r^2d\Omega_2=
\frac{1}{R^2(\tau)}\left(\frac{1+\gamma_0\rho/4}
{1-\gamma_0\rho/4}\right)^2
\left[d\tau^2-\frac{R^2(\tau)}{[1-\gamma_0^2\rho^2/16]^2}
\left(d\rho^2+\rho^2d\Omega_2\right)\right],
\label{Z58}
\end{eqnarray} 
we see that in the rest frame a topologically open RW cosmology with negative 3-curvature $K$ (viz. the one preferred in conformal cosmology \cite{Mannheim2006}) acts just like a universal linear potential with strength $\gamma_0=2(-K)^{1/2}$.

The inhomogeneous contribution is generated by the fourth integral in (\ref{Z49}). And in the region $r<r_{\rm clus}$ where $r_{\rm clus}$ is a typical cluster scale at which the global $f(r^{\prime})$ begins, this integral acts as a quadratic potential $-\kappa r^2$ with universal strength $\kappa=(1/6)\int_{r_{\rm clus}}^{\infty}dr^{\prime}r^{\prime}f(r^{\prime})$. Finally, adding the contributions of the global linear and quadratic potentials to the local (\ref{Z56}) yields for the total velocity and its asymptotic limit
\begin{equation}
v^2_{\rm TOT}(R)=v^2_{\rm LOC}(R)+\frac{\gamma_0 c^2R}{2}-\kappa c^2R^2,
\label{Z59}
\end{equation}                                 
\begin{equation}
v^2_{\rm TOT}(R)\rightarrow \frac{N^*\beta^*c^2}{R}+
\frac{N^*\gamma^*c^2R}{2}+\frac{\gamma_0c^2R}{2}-\kappa c^2R^2.
\label{Z60}
\end{equation} 
Despite the  fact that the galaxy-dependent $N^*$ is the only free parameter in (\ref{Z59}), using the values  
\begin{equation}
\beta^*=1.48\times10^5~{\rm cm},~\gamma^*=5.42\times 10^{-41} {\rm cm}^{-1},~\gamma_0=3.06\times 10^{-30} {\rm cm}^{-1},~\kappa=9.54\times 10^{-54} {\rm cm}^{-2}
\label{Z61}
\end{equation} 
Mannheim and O'Brien \cite{Mannheim2011b,Mannheim2012d,O'Brien2012} were able to fit the rotation curves of a large and varied sample of 138 spiral galaxies using only the visible galactic $N^*$ and no dark matter at all, viz. just one parameter per galaxy fits. In Fig. (1)  we show fits to 21 of the largest galaxies in the 138 galaxy sample. Since the $\kappa$-dependent term appears with a negative sign in (\ref{Z59}) and (\ref{Z60}), there would have to be an $N^*$-dependent bound on the size of a galaxy with $N^*$ stars since $v^2_{\rm TOT}(R)$ cannot be negative. To illustrate this effect we have included in Fig. (1) Mannheim and O'Brien's previously unpublished extended distance expectations for the three galaxies in the sample with the largest $N^*$ values. The novel prediction of an explicit end to galaxies provides a falsifiable signal for the conformal gravity theory.

That we are able to fit the data at all with a formula as constrained and universal as (\ref{Z59}) is very encouraging for the conformal theory, and especially so since, in addition to each visible galactic mass, dark matter fits to the same 138 galaxy sample require an additional 276 free parameters (two free parameters for each galactic dark matter halo). The reason why the fits do succeed is because across  the entire galaxy sample the value of $v^2_{\rm TOT}(R)/c^2R$ at the last detected data point of each galaxy is numerically found \cite{Mannheim2011b,Mannheim2012d,O'Brien2012} to not only be close to universal but to be close in magnitude to the universal $\gamma_0$. Such universality is a property of  the data themselves, and has yet to be accounted for by dark matter theory.

To conclude, we note that the fits not only provide support for the viability of conformal gravity at the macroscopic level per se, but in addition they provide evidence for the relevance to physics of an  underlying microscopic theory  that is based on a non-diagonalizable but $PT$-symmetric Hamiltonian.

{}

\begin{figure}[t]
\epsfig{file=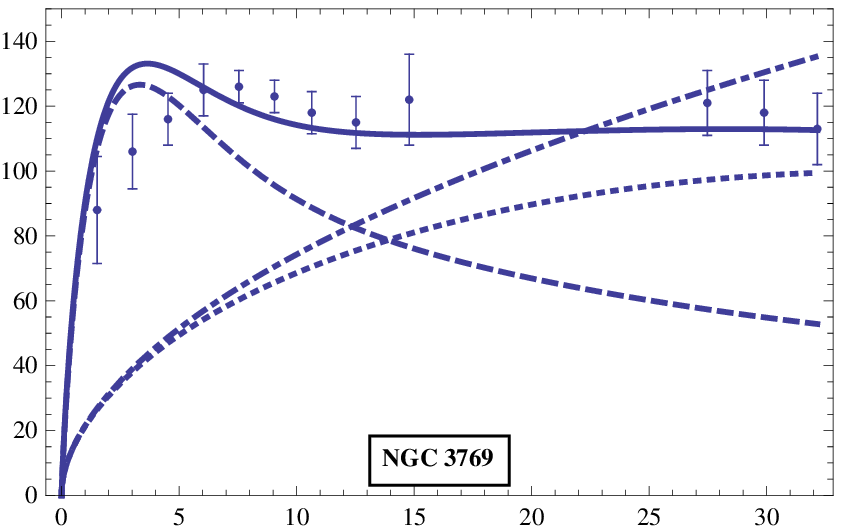,width=2.11in,height=1.2in}\qquad
\epsfig{file=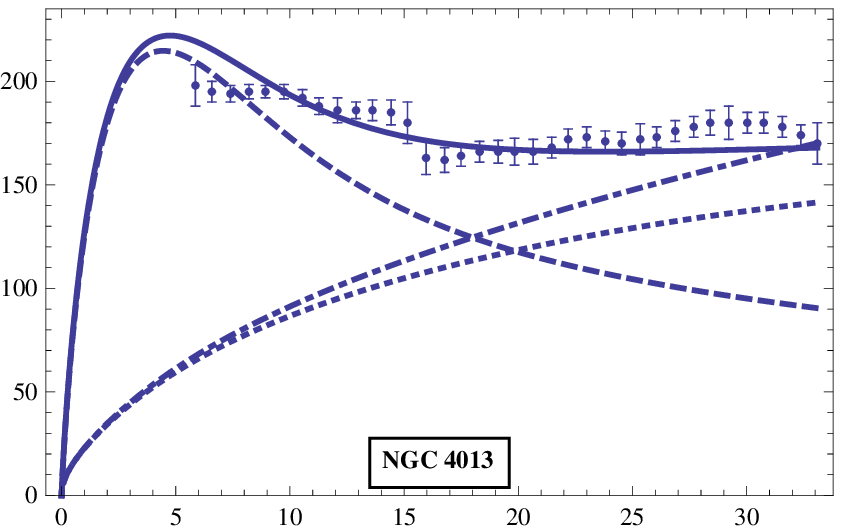,width=2.11in,height=1.2in}\qquad
\epsfig{file=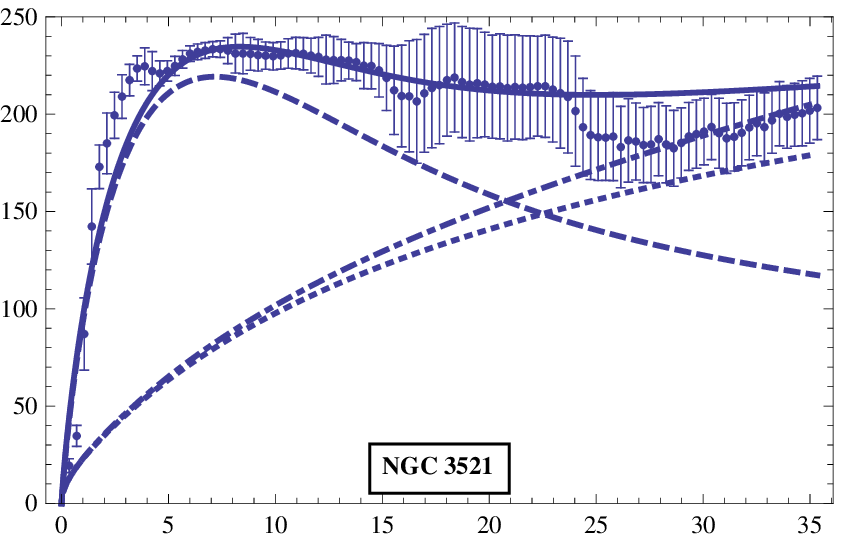,width=2.11in,height=1.2in}\\
\smallskip
\epsfig{file=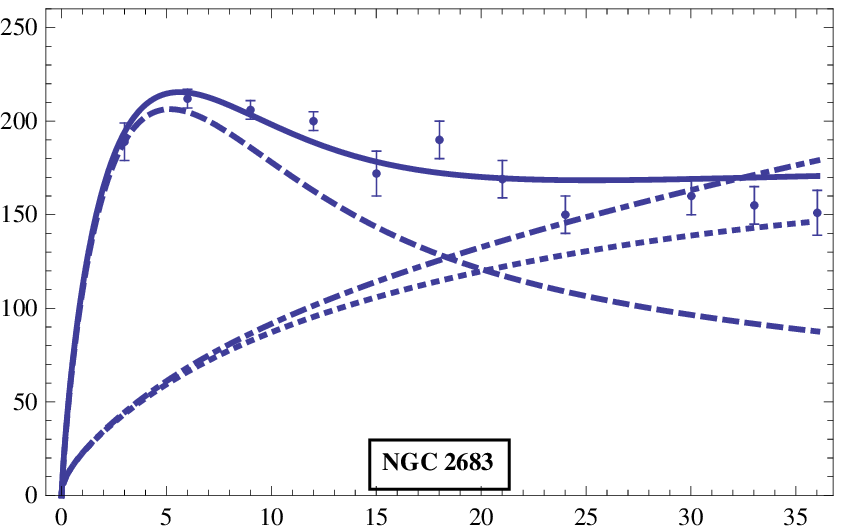,width=2.11in,height=1.2in}\qquad
\epsfig{file=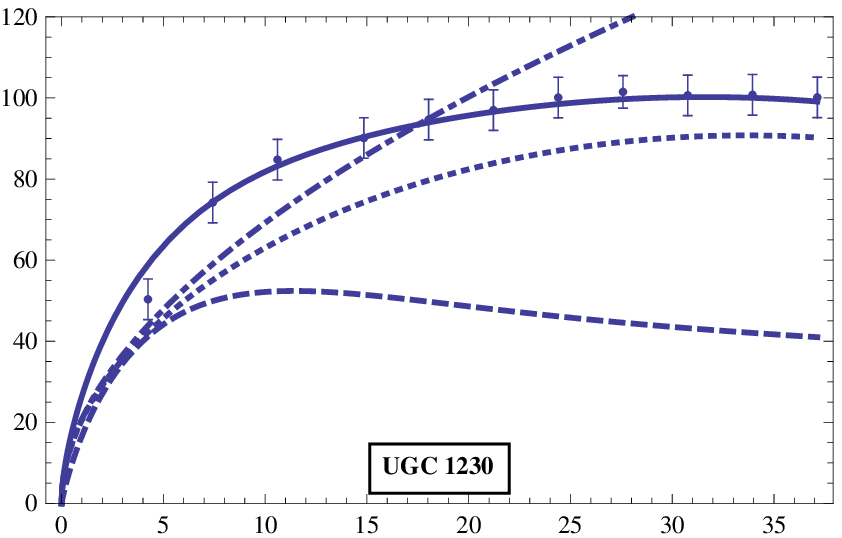,width=2.11in,height=1.2in}\qquad
\epsfig{file=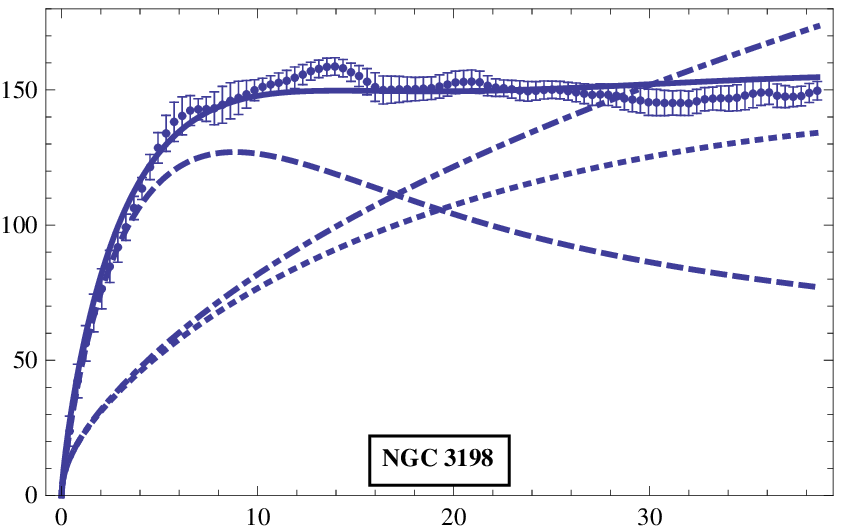,width=2.11in,height=1.2in}\\
\smallskip
\epsfig{file=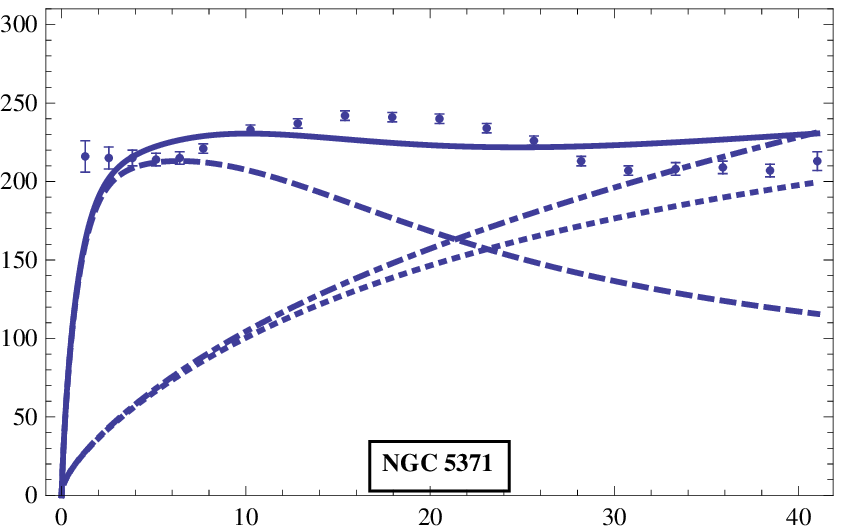,width=2.11in,height=1.2in}\qquad
\epsfig{file=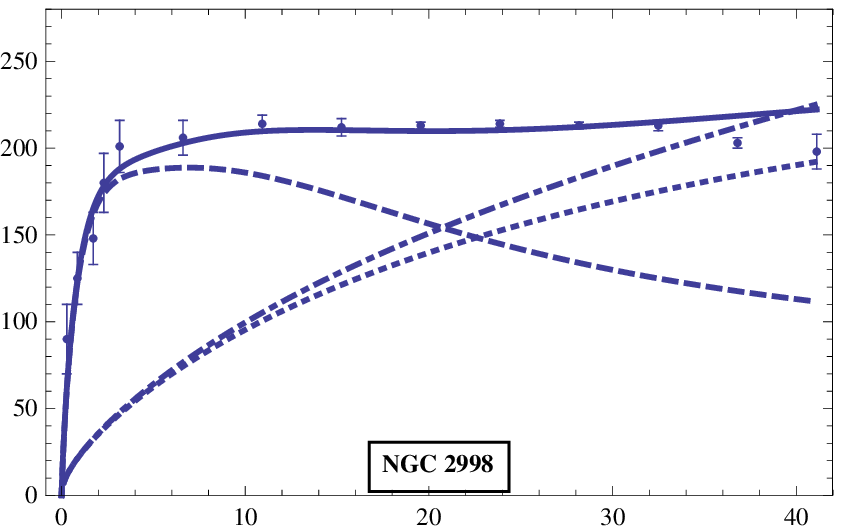,width=2.11in,height=1.2in}\qquad
\epsfig{file=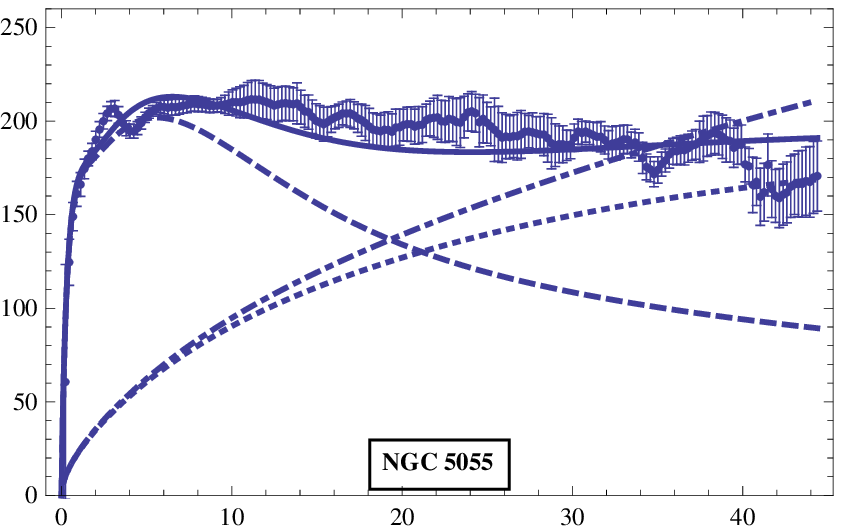,width=2.11in,height=1.2in}\\
\smallskip
\epsfig{file=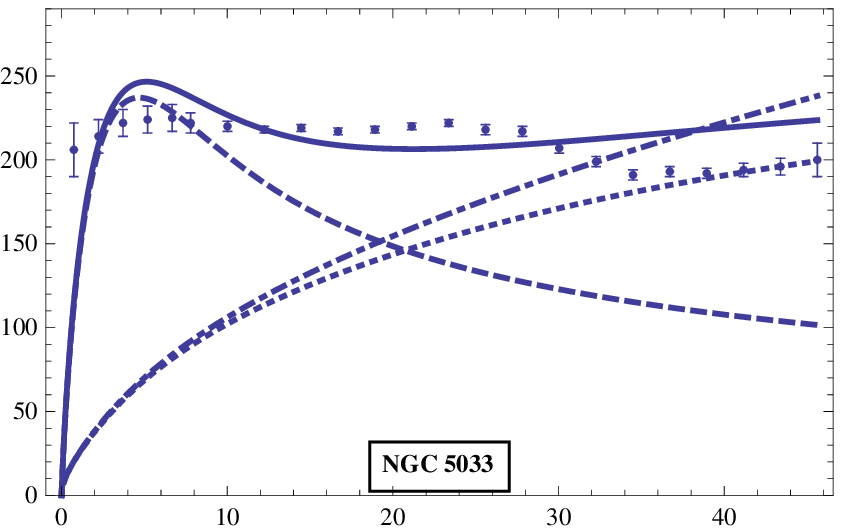,width=2.11in,height=1.2in}\qquad
\epsfig{file=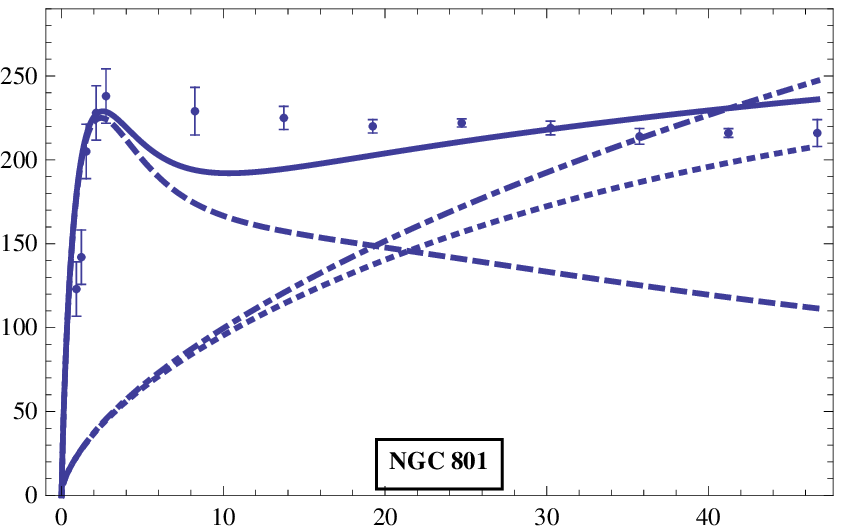,  width=2.11in,height=1.2in}\qquad
\epsfig{file=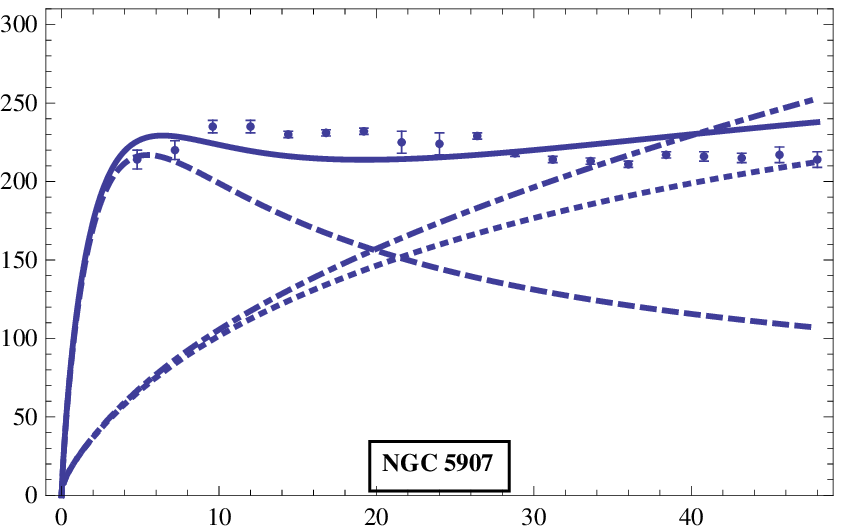,width=2.11in,height=1.2in}\\
\smallskip
\epsfig{file=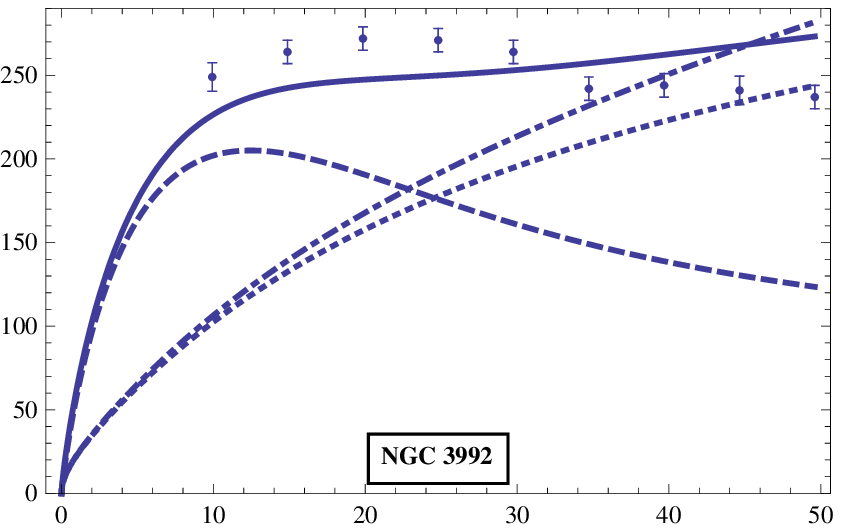,width=2.11in,height=1.2in}\qquad
\epsfig{file=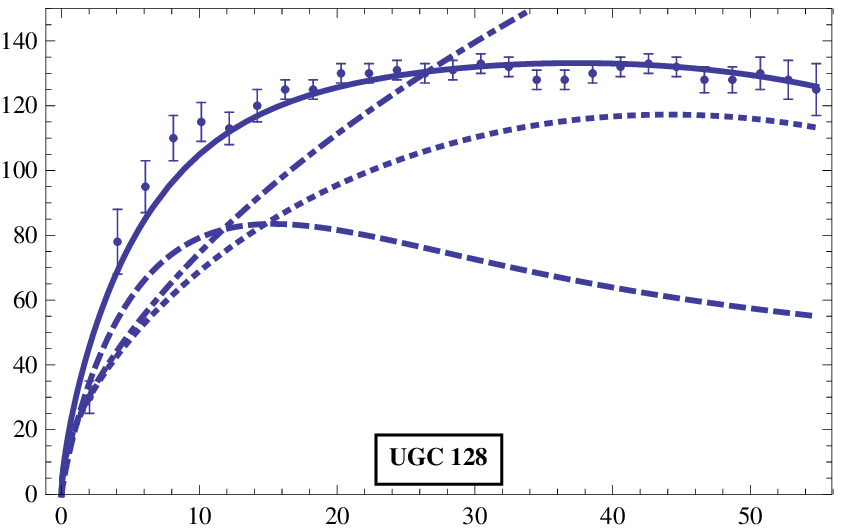,width=2.11in,height=1.2in}\qquad
\epsfig{file=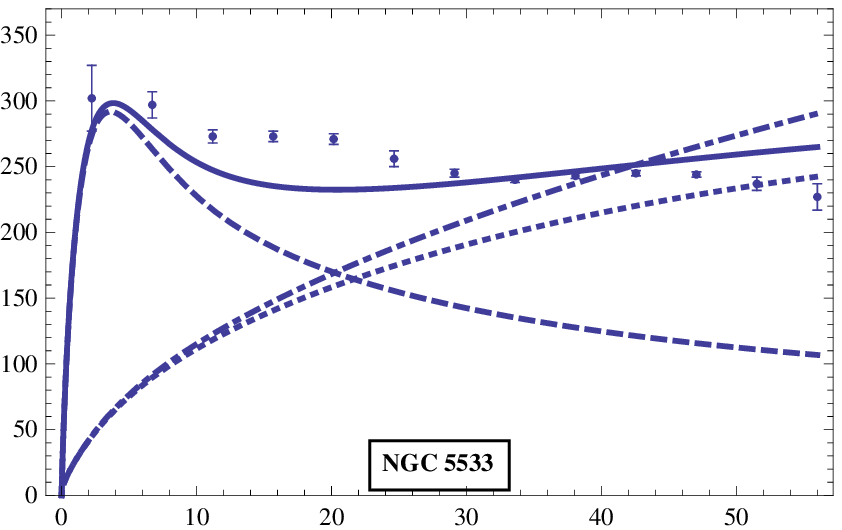,width=2.11in,height=1.2in}\\
\smallskip
\epsfig{file=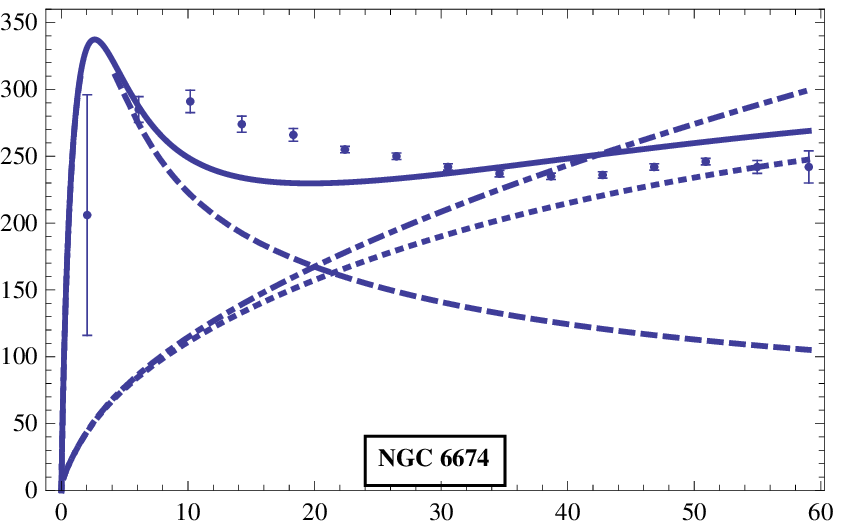,width=2.11in,height=1.2in}\qquad
\epsfig{file=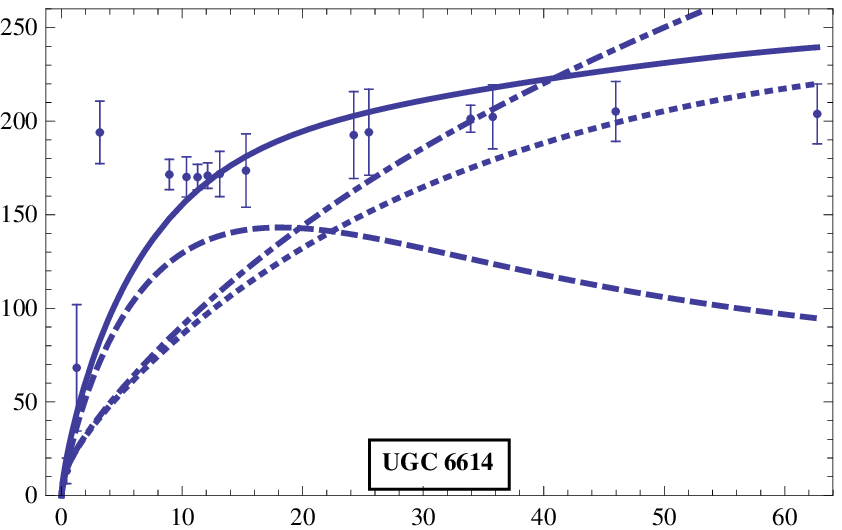,width=2.11in,height=1.2in}\qquad
\epsfig{file=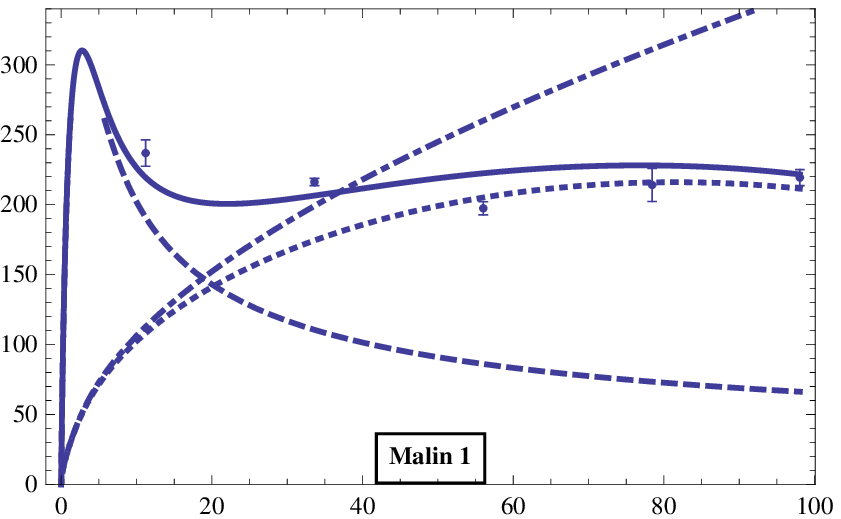,width=2.11in,height=1.2in}\\
\smallskip
\epsfig{file=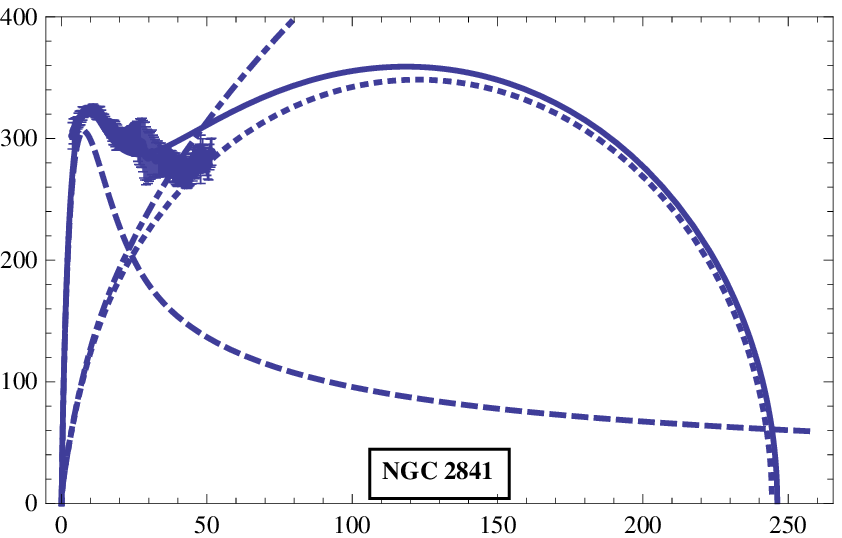,width=2.11in,height=1.2in}\qquad
\epsfig{file=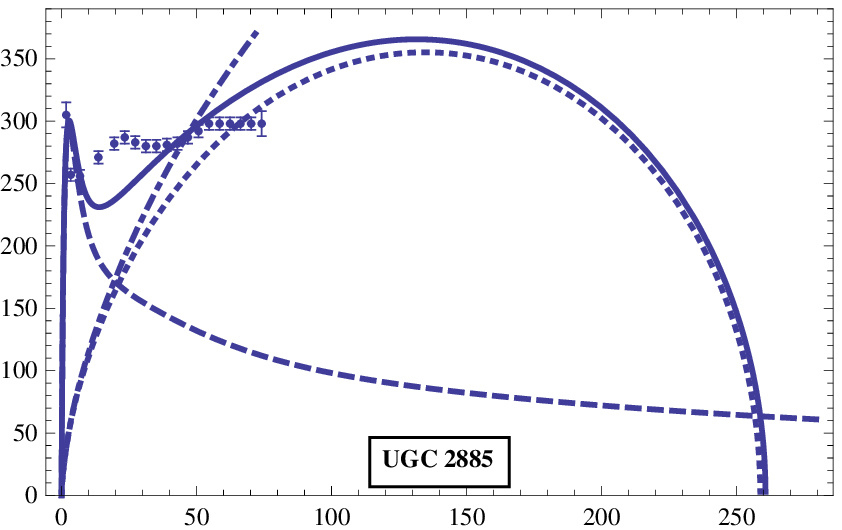, width=2.11in,height=1.2in}\qquad
\epsfig{file=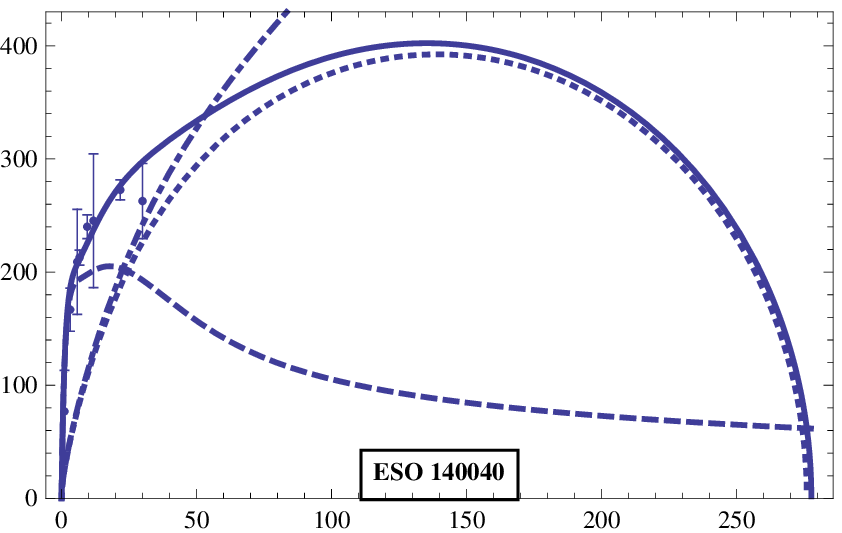,width=2.11in,height=1.2in}\\
\caption{Fitting to the rotational velocities of  21 large galaxies}
\label{Fig. (1)}
\end{figure}


\begin{thebibliography}{99}

\bibitem{Mannheim2000} P.~D.~Mannheim and A.~Davidson, \textit{Fourth order theories without ghosts}, January 2000. (arXiv:hep-th/0001115) 

\bibitem{Mannheim2005} P.~D.~Mannheim and A.~Davidson, Phys.~Rev.~A~\textbf{71}, 042110 (2005).  

\bibitem{Mannheim2007} P.~D.~Mannheim, ~Found.~Phys.~\textbf{37}, 532 (2007). 

\bibitem{Mannheim2006} P.~D.~Mannheim,~Prog.~Part.~Nucl.~Phys.~\textbf{ 56},~340~(2006). 

\bibitem{Mannheim2012a} P.~D.~Mannheim,~Found.~Phys.~\textbf{42}, 388 (2012).




\bibitem{Bender2008a}  C.~M.~Bender and P.~D.~Mannheim,~Phys.~Rev.~Lett.~\textbf{100},~110402 (2008).

\bibitem{Bender2008b} C.~M.~Bender and P.~D.~Mannheim,~Phys.~Rev.~D~\textbf{78},
025022 (2008).

\bibitem{Bender2008c} C.~M.~Bender and P.~D.~Mannheim, J.~Phys.~A~\textbf{41}, 304018 (2008).


\bibitem{Bender2007} C.~M.~Bender,~Rep.~Prog.~Phys.~\textbf{70}, 947 (2007).


\bibitem{Pais1950} A.~Pais and G.~E.~Uhlenbeck, Phys.~Rev.~\textbf{79}, 145 (1950).


\bibitem{Bender2010} C.~M.~Bender and P.~D.~Mannheim,~Phys.~Lett.~A~\textbf{374},
1616 (2010).



\bibitem{Mannheim2012b} P.~D.~Mannheim~\textit{$PT$ symmetry as a necessary and sufficient condition for unitary time evolution}, February 2012. (arXiv:0912.2635 [hep-th])

\bibitem{Mannheim2011a} P.~D.~Mannheim,~Gen.~Rel.~Gravit.~\textbf{43}, 703 (2011).

\bibitem{Mannheim1989} P.~D.~Mannheim and D.~Kazanas,~Ap.~J.~\textbf{342}, 635 (1989).

\bibitem{Mannheim1994} P.~D.~Mannheim and D.~Kazanas,~Gen.~Rel.~Gravit.~\textbf{26}, 337 (1994).

\bibitem{Mannheim2012c} P.~D.~Mannheim~\textit{Cosmological Perturbations in Conformal Gravity}, Phys.~Rev.~D in press. (arXiv:1109.4119 [gr-qc])


\bibitem{Mannheim1997} P.~D.~Mannheim,~Ap.~J.~\textbf{479}, 659 (1997).


\bibitem{Mannheim2011b} P.~D.~Mannheim and J.~G.~O'Brien,~Phys.~Rev.~Lett.~\textbf{106}, 121101 (2011).

\bibitem{Mannheim2012d} P.~D.~Mannheim and J.~G.~O'Brien, \textit{Fitting galactic rotation curves with conformal gravity and a global quadratic potential}, Phys.~Rev.~D in press. (arXiv:1011.3495 [astro-ph.CO])

\bibitem{O'Brien2012}  J.~G.~O'Brien and P.~D.~Mannheim,  MNRAS~\textbf{421}, 1273 (2012). 



\end{thebibliography}
\end{document}